# Research on Trends in Illegal Wildlife Trade based on Comprehensive Growth Dynamic Model

Run-Xuan Tang[1,*,#]

[1]School of Economics and Management, Beijing Jiaotong University, Beijing, China, 100044

*Corresponding author: 22241101@bjtu.edu,cn

#These authors contributed equally.

**Abstract:** This paper presents an innovative Comprehensive Growth Dynamic Model (CGDM). CGDM is designed to simulate the temporal evolution of an event, incorporating economic and social factors. CGDM is a regression of logistic regression, power law regression, and Gaussian perturbation term. CGDM is comprised of logistic regression, power law regression, and Gaussian perturbation term. CGDM can effectively forecast the temporal evolution of an event, incorporating economic and social factors. The illicit trade in wildlife has a deleterious impact on the ecological environment. In this paper, we employ CGDM to forecast the trajectory of illegal wildlife trade from 2024 to 2034 in China. The mean square error is utilized as the loss function. The model illuminates the future trajectory of illegal wildlife trade, with a minimum point occurring in 2027 and a maximum point occurring in 2029. The stability of contemporary society can be inferred. CGDM's robust and generalizable nature is also evident.

## 1. Introduction

The global wildlife trade, a complex phenomenon that transcends taxonomic and geographic boundaries, is a critical issue in conservation and biosecurity [1]. With millions of organisms traded annually, this activity threatens not only species survival but also poses health risks due to zoonotic disease transmission, as evidenced by the COVID-19 pandemic [2].

It is noteworthy that online platforms facilitate an unprecedented expansion of wildlife trade, enabling a wider reach and greater anonymity. This phenomenon has been documented in Asia [3] and Mexico [4]. The pandemic further highlighted the interface between wildlife trade and disease, prompting demand reduction campaigns and regulatory discussions [5]. However, effective interventions face challenges such as inadequate laws, monitoring limitations, and the lack of scientific evidence to support environmentalism [6]. This paper synthesizes the state-of-the-art literature on wildlife trade, focusing on illegal aspects, examining the roles of social media, consumer behaviors, and trader motivations, as well as responses to the pandemic [7].

A multidisciplinary approach is employed to analyze cases across Asia, Latin America, Africa, Europe, and Australia [8]. The efficacy of interventions, cultural nuances, enforcement gaps, and the convergence of organized crime are assessed [9].

This introduction draws upon a diverse range of literature [10] to explore trade dynamics, drivers, conservation implications, and human well-being. It emphasizes the necessity for nuanced strategies amidst evolving challenges.

## 2. The basic fundamental of CGDM

### 2.1 The structure of CGDM

The Comprehensive Growth Dynamic Model (CGDM) is a multifactor time series forecasting model. It incorporates a logistic growth component, a power-law social factor, an autoregressive component, and a Gaussian error term. This model comprehensively captures and forecasts the growth dynamics of complex economic and social phenomena. It not only takes into account the effects of economic and social factors, but also incorporates the autoregressive properties of the time series data itself. The mathematical expression is shown below.

$$Y(t) = T(t) + P(t) + A(t) + \varepsilon(t) \tag{1}$$

The minimization of the mean square error (MSE) loss function enables the model to enhance the precision and dependability of its forecasts. CGDM is a comprehensive and adaptable forecasting instrument for a diverse array of intricate time series data analysis.

## 2.2 Capturing the limits and dynamics of economic growth

In the Comprehensive Growth Dynamic Model (CGDM), economic factors (T(t)) are modeled by means of a logistic function. This is done to capture the growth problem with finite resources such as population, market size, etc. The mathematical expression is shown below.

$$T(t) = \frac{C}{1 + \exp(-k(t - m))} \tag{2}$$

The central strength of the logistic function is its ability to model a gradual slowdown in the rate of growth over time. The model is capable of demonstrating that the growth rate will eventually reach a plateau until it reaches a stable limit value. This property renders the Logistic model a valuable tool for comprehending and forecasting long-term growth trends.

C, which represents the capacity limit, is a parameter that determines the final boundary of growth. In the Logistic component of the CGDM, C signifies the maximum potential capacity for system growth. This implies that C is the highest level of growth that can be attained. In practice, C can correspond to market potential, environmental carrying capacity, or population limits, among others. It is therefore essential to accurately estimate C in order to be able to forecast and analyze growth trends.

Parameter k (growth rate) measures how fast or slow the growth rate is. The k parameter represents the growth rate in the Logistic part of the CGDM. It determines how fast or slow the growth rate will be. A higher value of k means that the system will approach its capacity limit faster. Therefore, accurate estimation of k is essential to capture the growth dynamics of economic factors.

Parameter m (growth saturation point) is a specific point on the logistic curve. It indicates the point in time when the growth rate reaches its maximum. Prior to this point, the growth rate increases gradually. Subsequent to this point, the growth rate begins to decline. This property enables CGDM to capture the critical period in the growth process.

In previous studies, Logistic models have been widely employed to predict and analyze long-term trends in global population growth and economic development.

## 2.3 Incorporating Time Series Autocorrelation

In the Comprehensive Growth Dynamic Model (CGDM), an autoregressive term (A(t)) is incorporated into the model. This is done to account for the autocorrelation of the time series data, i.e., the effect of past values on current values. The mathematical expression is shown below.

$$A(t) = \sum_{i=1}^{p} \omega_i \cdot A(t)_{t-i} \tag{4}$$

The autoregressive structure allows the model to capture and utilize historical information in the time series data. This enhances the accuracy of the forecasts.

$\omega_i$ (autoregressive weights) is a parameter that quantifies the effect of historical data on current values. In the autoregressive part of the CGDM, the $\omega_i$ parameter represents the weight of the impact of the historical values on the current values. Where $i$ ranges from 1 to $p$ and represents the number of historical periods considered by the model, these weights determine the relative importance of past observations on the current predicted values. This introduces a temporal dynamic feature into the model.

By employing $p$-order autoregression, the CGDM is able to model the dynamics of time series data in greater detail. This ensures that the model is able to take full advantage of the intrinsic linkages in the historical data when making predictions. The incorporation of this structure allows the CGDM to reflect not only the impact of economic and social factors. It is also able to take into account the autoregressive properties of the time series data itself. This enhances the precision and dependability of the forecasts.

**2.4 Capturing and Quantifying Forecast Uncertainty**

In the Comprehensive Growth Dynamic Model (CGDM), a Gaussian error term (ε(t)) is incorporated into the model. This is done to capture and quantify stochastic fluctuations that the model fails to explain. The mathematical expression is shown below.

$$\varepsilon(t) \sim \mathcal{N}(0, \sigma^2) \tag{5}$$

These fluctuations may originate from unpredictable market changes, policy influences, or other external factors. Their impact on the time series data is usually random.

$\sigma^2$ (error variance) is a parameter that measures the uncertainty of the forecast. In the CGDM, the $\sigma^2$ parameter represents the variance of the Gaussian error term. It is a key measure of the model's prediction uncertainty. A large value of $\sigma^2$ indicates a high level of uncertainty in the model prediction results. In contrast, smaller $\sigma^2$ values imply more reliable predictions. By estimating and monitoring $\sigma^2$, researchers and policy makers can better understand and assess the risk of model predictions.

The introduction of the Gaussian error term not only enhances the ability of CGDM to adapt to the complexity of the real world, but also provides a quantitative assessment of the uncertainty of model prediction results. This is essential for developing robust strategies and addressing potential risks. By combining economic, social, autoregressive, and Gaussian error terms, CGDM becomes a comprehensive and flexible forecasting tool for a wide range of complex time series data analyses.

**2.5 Loss Function**

The loss function plays a pivotal role in the Comprehensive Growth Dynamic Model (CGDM). It quantifies the discrepancy between the model's predicted values and the actual observed values. The mean square error (MSE) is a widely utilized loss function in regression problems. This is due to its ability to clearly illustrate the magnitude of the prediction error, facilitating mathematical processing and optimization. The formula for MSE is as follows.

$$MSE = \frac{1}{N} \sum_{t=1}^{N} \left(Y(t) - \hat{Y}(t)\right)^2 \tag{6}$$

The mean square error (MSE) is defined as the mean of the squared differences between the actual observations ($Y(t)$) and the model predictions ($\hat{Y}(t)$). The MSE is calculated for each data point and represents the average squared error between the observed and predicted values. The optimal parameters of the CGDM model are found by minimizing the MSE. These include $C$ (capacity limit), $k$ (growth rate), $m$ (growth saturation point), $\alpha$ (scale factor), $\beta$ (power law exponent), $\omega_i$ (autoregressive weights) and $\sigma^2$ (error variance). Thus, the predictions of the model are made as close as possible to the actual observations.

The minimization of the MSE is the central goal in the parameter optimization process of CGDM. This is usually achieved by optimization algorithms such as gradient descent. The model parameters can be effectively tuned to reduce prediction error and improve overall performance. CGDM is thus not only able to provide accurate forecasts but also to adapt to a variety of complex time series data analysis needs through its flexible framework structure and optimized parameters.

**2.6 Data preparation**

The data is presented in units of one individual wildlife animal. It is necessary to calculate imports and exports separately.

The data were obtained from the WILDLIFE SEIZURE DASHBOARD (https://wildlifedashboard.c4ads.org/) and IDEA (https://idea.usaid.gov/) as well as from COMAP (https://idea.usaid.gov/) websites for data acquisition and data cleaning. The data were obtained from the COMAP (https://www.comap.com/contests/mcm-icm) websites for data acquisition and data cleaning. This paper focuses on the illegal import and export of wildlife in China. The data were subjected to analysis and features were engineered. Finally, the model was input and computed.

## 3. Discussion of projection results

### 3.1 Training the model

The Python programming language is employed to implement CGMD.

Prior to the commencement of the iteration, the initial parameters must be set in accordance with the specifications outlined in the accompanying table 1.

Table 1 The initial parameters

| $C$ | $k$ | $m$ | $\alpha$ | $\beta$ | $\omega_1$ | $\omega_2$ | $\omega_3$ | $\sigma$ |
|---|---|---|---|---|---|---|---|---|
| 100 | 0.1 | 2 | 1 | 1 | r | r | r | 10 |

The random number $r_i$ ($i = 1, 2, 3$) is to be used in conjunction with the parameters $C, k,$ and $m$, which are employed to define the logistic function. This function describes the change in economic factors over time. The initial value of $C$ was set to $100$, as this value represents a large value that ensures the logistic function has a range of outputs. The initial values of $k$ and m were set to $0.1$ and $2$, respectively, as these values provide an idea of how the model changes.

The α and beta parameters are used to define the power law function, which describes how social factors change over time. The initial values of alpha and beta were selected because they represent the simplest forms of the power law function and can be utilized as a foundation for the model.

$\omega_i$, on the other hand, defines the autoregressive term that describes how past trade volumes affect current trade volumes. Three random values are selected as initial values because autoregressive models frequently necessitate optimization to identify the optimal weights.

σ This parameter is employed to define Gaussian noise, which is utilized to model the uncertainty in the model predictions. σ = 10 was selected as the initial value because it is a substantial value that ensures that the noise has sufficient range throughout the time horizon.

The parameters were optimized using gradient descent, and the model was tested with historical data. After training was completed, the parameters were as follows Table 2.

Table 2 The parameters after training

| $C$ | $k$ | $m$ | $\alpha$ | $\beta$ | $\omega_1$ | $\omega_2$ | $\omega_3$ | $\sigma$ |
|---|---|---|---|---|---|---|---|---|
| 109.9 | 0.7 | 2.4 | 41.0 | $1e-6$ | 0.2 | 0.1 | -0.2 | 10.0 |

The small value of β suggests that social events, including the 2019 New Crown Epidemic, did not cause a significant increase in illegal wildlife trade from 2015 to 2023. This implies that modern societies are stable and social events are less likely to cause significant problems.

The model was validated using the historical data. The results are shown in Fig 1.

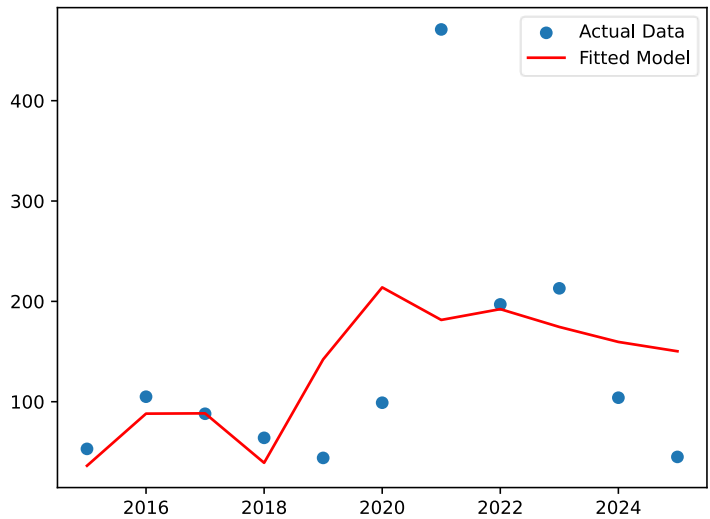

Fig 1. CGDM training process

Fig 1 depicts the CGDM training process. The Fig illustrates a satisfactory fit, with the 2021 value exhibiting clear anomalous behavior. The model effectively discards this data point, demonstrating its robust capability.

### 3.2 Analysis of experimental results

The CGDM is used to make projections the volume of illegal wildlife trade from 2024 to 2034. The results are shown in Fig 2.

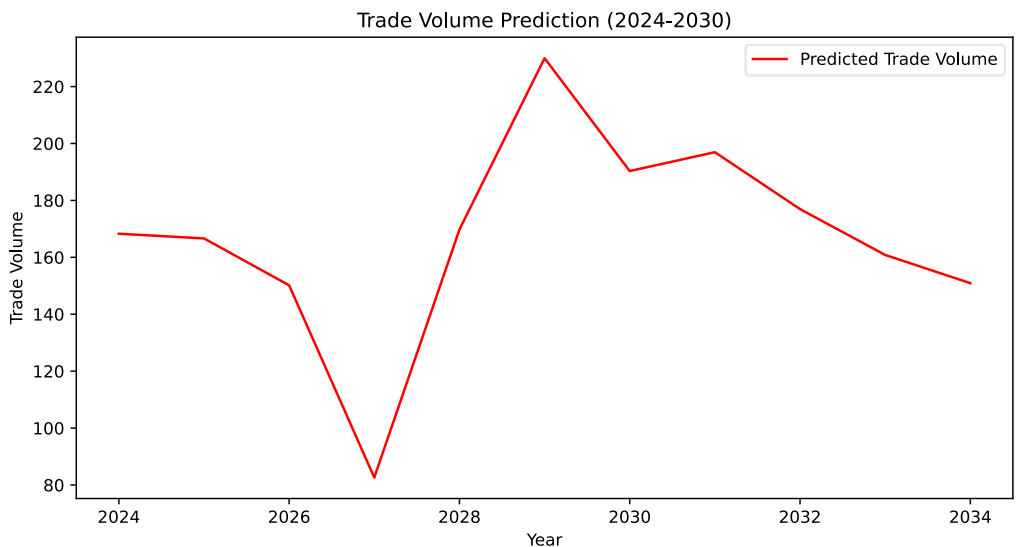

Fig 2. CGDM's Projections for the Future

Fig 2 demonstrate that the highest point of illegal trade in a decade will be in 2029. This indicates that control measures on illegal wildlife trade should be strengthened around 2029 to prevent the predicted trade peak from becoming a reality. Additionally, the model predicts that illegal wildlife trade will reach its lowest point in ten years in 2027. This suggests that certain factors may have prevented illegal trade from increasing in the years preceding 2029. Therefore, the measures taken in 2027 should be maintained. The overall trend exhibits a resemblance to historical data.

The model is based on economic and social factors, which indicates that the fluctuations in illegal trade are primarily driven by these two variables. The model employs autoregressive terms to capture the fluctuations in illegal trade, thereby enhancing the realism and accuracy of the forecasts.

The model is capable of forecasting fluctuations in the volume of illicit wildlife trade. Additionally, it can be employed to simulate and anticipate analogous occurrences.

**4. Conclusions**

This paper proposes the Comprehensive Growth Dynamics Model (CGDM), which models the impact of economic and social changes on specific events. The problem of illegal wildlife trade is addressed by applying CGDM to predict the illegal wildlife trade from 2024 to 2034. The mean square error is used as the loss function, and the model is optimized iteratively using the gradient descent method. The model is able to capture the trend of the data well and avoid overfitting. The results demonstrate the robustness and generalization ability of the model. The paper finds that modern societies are more stable and do not have social factors that cause higher-order power law growth. The results indicate that trade volumes will reach a low in 2027 and a high in 2029. This indicates the necessity of monitoring economic and social changes in the aforementioned years, as well as those preceding and following them. In order to effectively address the growth of the illegal wildlife trade, it is essential to implement effective policies and measures in 2027 and to make timely adjustments to strategies in 2029. The CGDM's simulation of the success of the illegal wildlife trade demonstrates its potential for future applications.